\documentclass[prb,twocolumn,showpacs]{revtex4}
\usepackage{graphicx}
\usepackage{amsmath}
\usepackage{amssymb}
\usepackage{dcolumn}
\usepackage{float}
\usepackage{bm}

\DeclareMathAlphabet{\bi}{OML}{cmm}{b}{it}

\def\be{\begin{equation}}
\def\ee{\end{equation}}
\def\bearr{\begin{eqnarray}}
\def\eearr{\end{eqnarray}}

\def\ra{\rangle}

\begin{document}
\title{Magnetotransport properties of two-dimensional fermionic systems with 
$k$-cubic Rashba spin-orbit interaction}
\bigskip
\author{Alestin Mawrie, Tutul Biswas and Tarun Kanti Ghosh}
\normalsize
\affiliation{Department of Physics, Indian Institute of Technology-Kanpur,
Kanpur-208 016, India}
\date{\today}
 
\begin{abstract}

The spin-orbit interaction in heavy hole gas formed at $p$-doped semiconductor
heterojunctions and electron gas at {\mbox SrTiO}${}_3$ 
surfaces is cubic in momentum. 
Here we report magnetotransport properties of 
k-cubic Rashba spin-orbit coupled two-dimensional fermionic systems.
We study longitudinal 
and Hall component of the resistivity tensor analytically as
well as numerically. The longitudinal resistivity shows beating 
pattern due to different Shubnikov-de Haas (SdH) oscillation frequencies $ f_{\pm} $ 
for spin-up and spin-down fermions. We propose empirical forms of 
$ f_{\pm} $ as exact expressions are not available, which are being 
used to find location of the beating nodes.
The beating nodes and the number of oscillations between any two
successive nodes obtained from exact numerical
results are in excellent agreement with those calculated from the
proposed empirical formula. 
In the Hall resistivity, an additional 
Hall plateau appears in between two conventional ones as spin-orbit 
coupling constant increases. 
The width of this additional plateau increases with spin-orbit coupling constant.

\end{abstract}

\pacs{73.61.Ey, 73.21.Fg, 71.43.-f, 71.70.Ej.}
\maketitle

%

\section{Introduction}

The spin degeneracy of a charge carrier in condensed matter systems 
is a result of spatial inversion symmetry and time-reversal 
symmetry together.
The spin degeneracy is lifted if one of the symmetries is absent.
The electric field normal to the interface of III-V semiconductor 
quantum well (such as {\mbox GaAs}/{\mbox AlGaAs} heterostructure) 
breaks the spatial inversion symmetry. The symmetry breaking electric 
field gives rise to spin-orbit interaction (SOI) which breaks the spin degeneracy 
even in absence of an external magnetic field. The dominant SOI
 in two-dimensional electron gas (2DEG) formed at the interface of 
{\mbox GaAs}/{\mbox AlGaAs} heterostructure is linear in momentum and 
it is known as Rashba SOI.\cite{rashba1,rashba2} 
Moreover, the Rashba SOI strength can be controlled with 
the help of an external bias.\cite{alpha1,alpha2,alpha3}
The SOI is essential to control and manipulate spin
degree of freedom of a charge carrier. After the proposal of spin field 
effect transistor by Datta and Das,\cite{datta} a large number of 
theoretical and experimental studies have been performed in the 
emerging field of spintronics \cite{cahay,sarma,acta} for the possibility of detecting pure 
spin current.
The Rashba SOI need not be always linear in
momentum. There are couple of systems where the Rashba SOI is
cubic in momentum. Two such systems are two-dimensional hole gas (2DHG)
\cite{2dh1,2dh2,2dh3, Winkler}
formed at p-doped {\mbox GaAs}/{\mbox AlGaAs} heterojunction and 2DEG 
at {\mbox SrTiO}${}_3$ surfaces.\cite{SrTi1,SrTi}
Very recently, it is reported in Ref. \cite{moriya} that $k^3$ SOI is dominating in
a two-dimensional hole gas formed in a strained \mbox{Ge}/\mbox{SiGe} quantum well.

In general, $4\times4$ Luttinger Hamiltonian\cite{luttin1,luttin2} 
describes 2DHG formed at $p$-doped semiconductor quantum well.
At very low temperature and low density only the lowest heavy hole (HH) 
sub-bands are occupied. The projection of $4 \times 4$ Luttinger Hamiltonian 
onto the HH states $\vert 3/2,\pm3/2\rangle$ leads to an effective $k$-cubic\cite{2dh2,2dh3} 
Rashba SOI.
This SOI opens up a gap $\Delta_{\rm so}=2\alpha k_F^3$ 
(where $\alpha$ is the Rashba spin-orbit coupling constant and 
$k_F$ is the Fermi wave vector) between two spin-split heavy hole 
sub-bands. 
This spin splitting has been observed experimentally\cite{Grbic} for holes 
in C-doped p-type GaAs/AlGaAs quantum well and the value of $\Delta_{\rm so}$ 
is also extracted by analyzing the beating pattern in the SdH oscillations as 
well as by using weak anti-localization method. 
In recent past a number of investigations have been performed 
to explore various properties of 2DHG like effective mass,\cite{mass1, mass2, mass3} 
effective g factor,\cite{lande} spin polarization,\cite{polar}
spin rotation\cite{rotation} etc. 
Most importantly, spin Hall effect\cite{2dh3, she1, she2, she3, she4, she5, she6}
in 2DHG has been studied experimentally as well as theoretically.

On the other hand, three 3d orbitals ($t_{2g}: d_{xy}, d_{xz}, d_{yz}$) of 
Ti ion form the conduction band\cite{SrTi2, SrTi3} of {\mbox SrTiO}${}_3$ crystal.
The 3d orbitals at the surface are confined in the z-direction which
is normal to the interface. As a result it yields a 2DEG. 
The lowest energy states of the bulk {\mbox SrTiO}${}_3$ are fourfold 
degenerate bands, corresponds to the states $ | 3/2,\pm 3/2 \ra $ and
$ 3/2,\pm 1/2 \ra $. 
Various studies\cite{SrTi4, SrTi5} suggest that the confinement along the z direction
lifts the $\Gamma $-point degeneracy between the $d_{xy} $ band and        
$ d_{xz}, d_{yz} $ bands. Within the effective tight-binding Hamiltonian
for $t_{2g} $ bands of {\mbox SrTiO}${}_3$ surfaces, the Rashba
SOI is cubic in momentum. There is an experimental evidence\cite{SrTi1} of 
$k$-cubic Rashba SOI on {\mbox SrTiO}${}_3$ surfaces.

In this work we study magnetotransport coefficients of fermions with 
$k$-cubic Rashba spin-orbit interaction. The longitudinal conductivity 
which arises entirely due to the collision or hopping process exhibits 
beating pattern. 
We provide two empirical frequencies $f_\pm$ of quantum oscillations 
of spin-up and spin-down fermions, which are responsible 
for the beating pattern. 
At higher values of magnetic field the beating pattern is replaced 
by resistivity peak. 
As $\alpha$ increases, the peak in the resistivity splits into two asymmetric 
ones. On the other hand, Hall conductivity shows the conventional plateau 
structure. With the increase of $\alpha$ an additional plateau arises between 
two conventional ones. The width of this additional 
plateau increases with $\alpha$.

This paper is organized as follows. In section II, we present the 
basic informations about the $k$-cubic spin-orbit coupled fermions. 
In section III, analytical calculations for the
different transport coefficients are given. Numerical results and discussion 
are presented in section IV. We summarize our results in section V.

\section{basic informations}
The Hamiltonian for a fermion with k-cubic spin-orbit interaction
in presence of a magnetic field ${\bf B} = B \hat z$ is given by\cite{Liu, Zarea} 

\begin{eqnarray}\label{hamil2}
H & = & \frac{{\bf \Pi}^2}{2m^\ast} + 
\frac{i\alpha}{2\hbar^3}\big(\Pi_{-}^3\sigma_{+} - \Pi_{+}^3 \sigma_{-} \big) - 
\frac{3}{2}g^\ast{\mu_{B}}{\boldsymbol{\sigma}} \cdot {\bf{B}},
\end{eqnarray}
where ${\bf \Pi}={\bf p}- e{\bf A}$ with ${\bf A}$ is the vector potential,
$m^\ast$ is the effective mass of the fermion and $\alpha$ is Rashba spin-orbit 
coupling constant. Also, $p_\pm = p_x \pm i p_y$,  
$ \sigma_\pm = \sigma_x \pm i \sigma_y$ 
with $\sigma_i$'s are the Pauli matrices, $g^\ast$ 
is the effective Lande-g factor and $\mu_{B}$ is the Bohr magneton.

Using the Landau gauge ${\bf{A}}=(0,xB,0)$, the Hamiltonian given by 
Eq. (\ref{hamil2}) commutes with $p_y$ i.e. $k_y$ 
is a good quantum number in this case.
The energy eigen value for $n \geq 3 $ is given by
\begin{equation}
E_n^{\lambda}=\hbar\omega_c\Big[n-1+
\lambda\sqrt{\tilde{E}_{n\alpha}^2+\tilde{E}_0^2}\Big],
\end{equation}
where $\lambda=\pm$, $\tilde{E}_0=3/2 - \chi$ with 
$ \chi=3g^\ast m^\ast/(4m_0)$, 
$\tilde{E}_{n\alpha}=\tilde{\alpha}\sqrt{8n(n-1)(n-2)}$.
Here $\tilde{\alpha}$ is defined as $\tilde{\alpha}=l_{\alpha}/l_c$
with $l_{\alpha}=m^*\alpha/\hbar^2$ and $l_c=\sqrt{\hbar/(eB)}$ is 
the magnetic length.
The corresponding eigenstates for positive and 
negative branches are given by
\begin{eqnarray} 
\psi_{n,k_y}^+(x,y)=\frac{e^{ik_yy}}{\sqrt{L_yA_n}} 
\begin{pmatrix} 
\phi_{n} ( X )
\\ D_n\phi_{n-3} ( X )
\end{pmatrix}
\end{eqnarray}
and
\begin{eqnarray} 
\psi_{n,k_y}^-(x,y)=\frac{e^{ik_yy}}{\sqrt{L_yA_n}} 
\begin{pmatrix} -D_n\phi_{n} (X)
\\ \phi_{n-3}( X ) \end{pmatrix},
\end{eqnarray}
where $L_y$ is the system length along $y$-direction, 
$ X = x - x_c $ with $x_c=k_yl_c^2$ 
and $A_n=1+D_n^2$ with 
$D_n=\tilde{E}_{n\alpha}/\Big(\tilde{E}_0 + 
\sqrt{\tilde{E}_0^2+\tilde{E}_{n\alpha}^2}\Big)$.
Here $\phi_n(x)$ is the oscillator wave function of order $n$.

For $n<3$ there is only one branch ($+$ branch). In this case 
the eigenvalues and eigen functions are given by
\begin{equation}
E_n=\hbar\omega _c (n + 1/2 - \chi )
\end{equation} 
and
\begin{eqnarray} 
\psi_{n,k_y}(x,y)=\frac{e^{ik_yy}}{\sqrt{L_y}}\phi_n 
( X )\begin{pmatrix} 1
\\ 0 \end{pmatrix}.
\end{eqnarray}
The derivation of the energy spectrum and the corresponding eigenfunctions are
given in Appendix A.

\section{Derivation of magnetotransport Coefficients}

In this section we calculate both the longitudinal and transverse 
components of conductivity tensor using Kubo formula.\cite{kubo} 
The longitudinal conductivity contains diffusive and collisional 
contribution. In presence of a perpendicular magnetic field the 
diagonal matrix elements of velocity operator become zero which 
in turn causes the vanishing of diffusive conductivity. So 
the longitudinal conductivity is solely due to the collisional 
contribution.

{\bf{Collisional conductivity}}: 
At low temperature, one can safely assume that fermions
are elastically scattered by charged impurities distributed uniformly
over the system. The expression for
collisional conductivity is given by \cite{Van, Carol, vasilo, Peet, wang}
\begin{eqnarray}\label{coll}
\sigma ^{\rm coll}_{xx}&=&\frac{\beta e^2}{\Omega}\sum\limits_{\xi,\xi^\prime}
f(E_\xi)\{1-f(E_{\xi^\prime})\} W_{\xi \xi^\prime}(x^{\xi}-x^{\xi^\prime})^2.
\end{eqnarray}
Here $\vert\xi\rangle=\vert n,k_y,\lambda\rangle$ defines a set of 
all quantum numbers, $\Omega$ is the surface area of the two-dimensional system, 
$f(E_\xi)=1/(\exp((E_\xi-\mu)\beta)+1)$ is the
Fermi distribution function with $\beta=1/(k_BT)$ and 
$x^\xi=\langle \xi\vert x\vert\xi\rangle=k_yl_c^2 $ is the expectation 
value of the $x$ component of the position operator.
Finally, the transition
probability between two states $|\xi\rangle$ and $|\xi^\prime\rangle$ is 
given by
\begin{eqnarray}\label{trans_prob}
W_{\xi \xi^{\prime}}&=&\frac{2\pi n_{\rm im}}{\hbar\Omega }
\sum\limits_{q}\vert U({\bf q})\vert^2\vert F_{\xi,\xi^\prime}\vert^2
\delta (E_{\xi}-E_{\xi^\prime}),
\end{eqnarray}
where $n_{\rm im}$ is the impurity density and 
$U(q)=e^2/(2\epsilon_0 \epsilon(q^2+k^2_s)^{1/2})$ 
is the Fourier transform of the screened Yukawa-type impurity potential 
$ U(r)=e^2e^{-k_sr}/(4\pi\epsilon_0 \epsilon r)$ with
$\epsilon$, $\epsilon_0$ and $k_s$ as the dielectric constant of the medium,
the vacuum permittivity and the screened wave vector, respectively.
The term $F_{\xi,\xi^\prime}=\langle{\xi}\vert e^{i{\bf q} 
\cdot{\bf r}}\vert{\xi^\prime}\rangle$ is called form factor whose
complete expression is given in Appendix B.
Since the term $F_{\xi,\xi^\prime}$ is proportional to 
$\delta_{k_y^\prime,k_y+q_y}$, the summation over 
$k_y^\prime$ in Eq. (\ref{coll}) can be easily evaluated with the 
replacement of $k_y^\prime$ by $k_y+q_y$ and we have 
$(x^\xi-x^{\xi^\prime})=q_y^2l_c^4=q^2l_c^4\sin^2\phi$. 
The delta function in Eq. (\ref{trans_prob}), 
$\delta (E_{\xi}-E_{\xi^\prime}) = 
\delta (E_n^\lambda-E_{n^\prime}^{\lambda^\prime})$, 
ensures the possibilities of only intra-branch and 
intra-level scattering i.e. $n^\prime=n$ and $\lambda^\prime=\lambda$. 
Again $\delta (E_n^\lambda-E_{n^\prime}^{\lambda^\prime})$ can be written
in its usual Lorentzian representation i.e. 
$\delta (E_n^\lambda-E_{n^\prime}^{\lambda^\prime})
=(1/\pi)\Gamma/[(E_n^\lambda-E_{n^\prime}^{\lambda^\prime})^2 + 
\Gamma^2]$  with $\Gamma$ is the impurity induced Landau level broadening. 
We also have $\sum_{k_y}\rightarrow \Omega/(2\pi l_c^2)$ and
$\sum_{{\bf q}}\rightarrow (\Omega/(2\pi)^2)\int qdqd\phi$. So by 
inserting Eq. (\ref{trans_prob}) into Eq. (\ref{coll}) and after doing all 
the summations one can obtain the following expression of the collisional 
conductivity:

\begin{eqnarray}\label{coll1}
\sigma_{xx}^{\rm coll}&=&\frac{e^2}{h}\frac{\beta n_il_c^2U_0^2}{2\pi\Gamma}
\sum_{n,\lambda}f(E_n^\lambda)\{1-f(E_n^\lambda)\}\nonumber\\
&\times&\int dqq^3\vert F_{nn}^\lambda(q)\vert^2.
\end{eqnarray}
In deriving Eq. (\ref{coll1}) we have used the following approximation
$\vert U(q) \vert \simeq U_0=e^2/(2\epsilon_0\epsilon k_s)$ since $q\ll k_s$. Now 
using the fact $n_{\rm im} U_0^2 \sim (\Gamma l_c)^2/(4\pi)$, we finally have

\begin{eqnarray}\label{coll2}
\sigma_{xx}^{\rm coll}=\frac{e^2}{h}\frac{\beta \Gamma}{4\pi^2}
\sum_{n,\lambda}f(E_n^\lambda)\{1-f(E_n^\lambda)\}I_n^\lambda,
\end{eqnarray}
where $I_n^\lambda=\int_0^{\infty} udu \vert 
F_{nn}^\lambda\vert^2$ with $u=q^2l_c^2/2$.
It is straightforward to evaluate the expression of $I_n^\lambda$ as
given by $ I_{n<3} = 2n+1 $ and 
$I_{n\geq 3}^\lambda = \{(2n-2-\lambda 3)(D_n^4+1)+\lambda 6\}/A_n^2$.

{\bf Hall conductivity:}
The expression for the Hall conductivity  $\sigma_{yx}$ is given by \cite{vasilo, Peet, wang}
\begin{eqnarray}\label{hall_con}
\sigma_{yx} & = & \frac{i\hbar e^2}{\Omega }\sum \limits_{\xi \neq \xi^{\prime}}
\langle \xi|v_y|\xi^{\prime} \rangle
\langle \xi^{\prime}|v_x|\xi \rangle\frac{f(E_{\xi}) - 
f(E_{\xi^{\prime}})}{(E_{\xi}-E_{\xi}^{\prime})^2}.
\end{eqnarray}

The matrix elements of the components of the velocity operator are given 
in Appendix C. By virtue of the Kronecker delta symbols in 
Eqs. (\ref{vellx}-\ref{velly}), it is confirmed that the 
transitions are allowed only between the adjacent Landau 
levels $n^\prime = n \pm 1 $. We mention here that inter-branch 
transitions also possible along with the intra-branch 
scattering. So the summation in Eq. (\ref{hall_con}) can be 
split into four terms as 
$\sigma_{yx} = \sigma_{yx}^{++} + \sigma_{yx}^{--} + 
\sigma_{yx}^{+-} + \sigma_{yx}^{-+}$, where the first two terms 
correspond to the intra-branch transition and the last two terms
correspond to the inter-branch transition.
Setting $\tilde{E}_n^\lambda=E_n^\lambda/(\hbar\omega_c)$,
the total Hall conductivity is given by

\begin{eqnarray}\label{HT}
\sigma_{yx}  & = & \frac{e^2}{h} 
\bigg[\sum_{n=0}^{1}(n+1)\{f(E_n)-f(E_{n+1})\}\nonumber\\
&+&\sum_{\lambda}\frac{f(E_2)-f(E^\lambda_{3})}{(\tilde{E}_2-\tilde{E}_{3\lambda})^2}
{C^\lambda_{23}}\nonumber\\
&+&\sum_{\lambda, n=3}^{\infty} 
\Big\{{C_n^{\lambda}}^2\frac{f(E_n^\lambda)-f(E_{n+1}^\lambda)}
{(\tilde{E}_n^{\lambda}-\tilde{E}_{n+1}^{\lambda})^2}\nonumber\\
&+&{C_n^{\prime\lambda}}^2\frac{f(E_n^{-\lambda})-f(E_{n+1}^{\lambda})}
{(\tilde{E}_n^{-\lambda}-\tilde{E}_{n+1}^{\lambda})^2}\Big\}\bigg],
\end{eqnarray}
where $C^+_{23}=(\sqrt{3/2}+6\sqrt{2}\tilde{\alpha}D_3)^2$, $C^-_{23}=(-D_3\sqrt{3/2}+6\sqrt{2}\tilde{\alpha})^2$,
 $C_n^+=B_n$, $C_n^-=K_n$, 
$C_n^{\prime+}=B_n^\prime$ and $C_n^{\prime-}=K_n^\prime$ 
are given in Appendix B.

\section{Numerical Results and Discussion}
In this section we shall present numerical results of 
Eqs. (\ref{coll2}) and (\ref{HT}).
Typical system parameters of p-type GaAs/AlGaAs heterostructure and
{\mbox SrTiO}${}_3$ materials are
summarized here. 
For 2DHG, $n_f = 2.03 \times10^{15}$ m$^{-2}$ is the charge carrier density, 
$m^\ast=0.41m_0$ with $m_0$ as the free electron mass and $g^\ast=6.5$.
For {\mbox SrTiO}${}_3$ materials \cite{SrTi1,Srmass},
$n_f=2.4-4.7\times10^{16}$ m$^{-2}$, $m^\ast=1.5 m_0$ and $g^\ast = 2 $.
The Landau level broadening is assumed to be $\Gamma=0.01$ meV.
Note that $ \Gamma \ll \Delta_{\rm so} \ll \hbar \omega_c $ \cite{fogler} so that
$ \Gamma $ does not blurred the discrete spectrum completely.
For various plots, we adopt the parameters of GaAs/AlGaAs heterostructure.
 
\begin{figure}[h]
\begin{center}\leavevmode
\includegraphics[width=93mm,height=70mm]{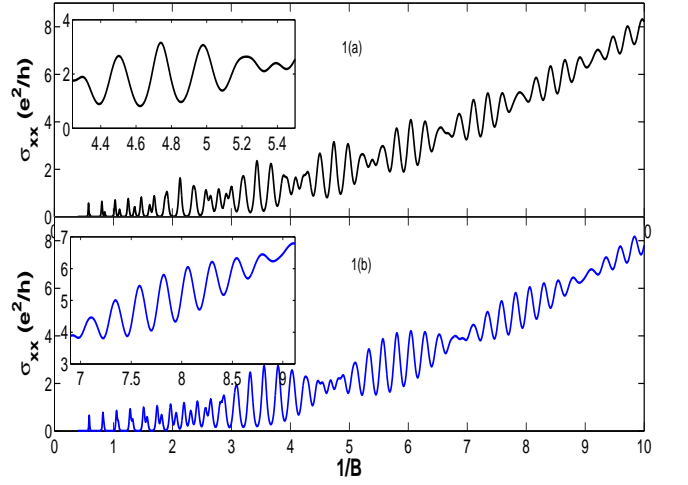}
\caption{(color online) Plot of $\sigma_{xx}$ as a function of $1/B$.
The upper panel (1(a)) is for $\alpha=0.080$ eVnm$^3$
and the lower panel (1(b)) is for $\alpha=0.048$ eVnm$^3$.
The oscillations between 3$^{rd}$ and 
4$^{th}$ nodes are shown in insets.}
\label{Fig1}
\end{center}
\end{figure}

Earlier the spin-splitting and hence the value of $\alpha$ was 
determined from the difference between the two spin-split heavy 
hole subband densities.\cite{Winkler, Grbic} The population densities in 
the two spin-split sub-bands are measured by analyzing the SdH oscillation 
frequencies \cite{Yuan}. In the present study we give
an alternative treatment for determining $\alpha$ by simply counting 
the number of oscillations between two beating nodes.

In Fig. 1 we show the variation of $\sigma_{xx}$ with the 
inverse magnetic field for two different values of $\alpha$:
$\alpha=0.08$ eVnm$^3$ and  $\alpha=0.048$ eVnm$^3$.
Figure 1 clearly shows regular beating pattern formation in $\sigma_{xx} $. 
Unlike the case of 2DEG with Rashba SOI,\cite{Firoz}
it is not possible to get an analytical expression of the 
density of states of Landau levels for two-dimensional fermions with $k$-cubic
Rashba SOI. It hinders to have closed-form 
analytical expression of $\sigma_{xx}$.
At the same time, exact positions of the nodes would help us to 
determine the value of $\alpha$. To obtain the exact locations of the
nodes as shown in Fig. 1, we model $\sigma_{xx}$ as
\begin{eqnarray}
\sigma_{xx}\propto \cos \left( 2\pi f_a/B \right)
\cos \left( 2\pi f_d/B \right),
\end{eqnarray}
where $f_a=(f_+ + f_-)/2$ and $f_d = (f_+-f_-)/2$ with $f_{\pm} $ are 
the SdH oscillation frequencies for spin-up and spin-down fermions.

Careful observations reveal that the SdH oscillation frequencies
for spin-up and spin-down electron in spin-orbit coupled 2DEG are
directly related to the spin-split Landau levels.
The approximate SdH oscillation frequencies can be obtained by
setting $n=n_F $, where $n_F$ is the Landau level quantum number corresponds
to the Fermi energy $E_F$, in the spin-split Landau levels. 
Using the same analogy for $k$-cubic spin-orbit coupled systems, 
we propose $f_\pm$ will have the following forms
 \begin{eqnarray}\label{freq}
 f_{\pm}=\frac{m^*}{\hbar e}\bigg(E_{F}^0\mp 
 \sqrt{\frac{8\alpha^2 (m^* E_{F}^0)^3}{\hbar^6}+E_0^2}\bigg),
 \end{eqnarray}
where $ E_{F}^0 = (\hbar k_{F}^0)^2/(2m^*) $ with 
$ k_{F}^0 = \sqrt{2\pi n_f}$.

Now the beating nodes are simply given by $2f_d/B_j=(j+1/2)$ with 
$j=1,2,...$. Using Eq. (\ref{freq}) we obtain the locations of the
nodes at
\begin{eqnarray}\label{beat}
\frac{1}{B_j}=\frac{e\hbar}{8\pi\alpha n_f m^*} 
\sqrt{\frac{(2j+1)^2-4(3-2\chi)^2}{2\pi n_f}}.
\end{eqnarray}
The number of oscillations between any two successive 
nodes is given by 
\begin{eqnarray}\label{Nosc}
N_{\rm osc}=f_a \Delta \left(\frac{1}{B}\right)
 =\left(\frac{m^*E_{F}^0}{\hbar e}\right) 
\left(\frac{1}{B_j}-\frac{1}{B_{j+1}}\right).
\end{eqnarray}
It is straightforward to find the expression for $\alpha$ by
solving Eq. (\ref{beat}) and (\ref{Nosc}) as
\begin{eqnarray}\label{alpha}
 \alpha=A\Big(\sqrt{(2j+3)^2-C^2}-\sqrt{(2j+1)^2-C^2}\Big),
\end{eqnarray}
where $A=\hbar^3/(N_{\rm osc}\sqrt{128m^{\ast3}E_{F}^0})$
and $C=2(3-2\chi)$.

Let us now check whether the results for the locations of nodes and
for the number of oscillations $N_{\rm osc} $, given by Eqs. (\ref{beat})
and (\ref{Nosc}),
obtained based on empirical expressions for $f_{\pm} $ match with
the exact numerical results given in Fig. 1.   
The positions of nodes calculated 
from both approach are summarized in table I. 
Moreover, the inset of Fig. 1 shows the number of oscillations between 
two successive beating nodes.
As seen from the inset there are $5$ and $9$ oscillations between 
$3^{rd}$ and $4^{th}$ nodes in Fig. 1(a) and Fig.1(b), respectively,
which are same as calculated from Eq. (\ref{Nosc}).
Again, 
the calculated strength of Rashba SOI ($\alpha$) taking $j=3$ in Eq.(\ref{alpha})
are $\alpha=0.0799$ eVnm$^{3}$ and $\alpha=0.0473$ eVnm$^{3}$, respectively. 
Thus the numerical and approximate results are in excellent agreement.

\begin{figure}[t]
\begin{center}\leavevmode
\includegraphics[width=95mm,height=70mm]{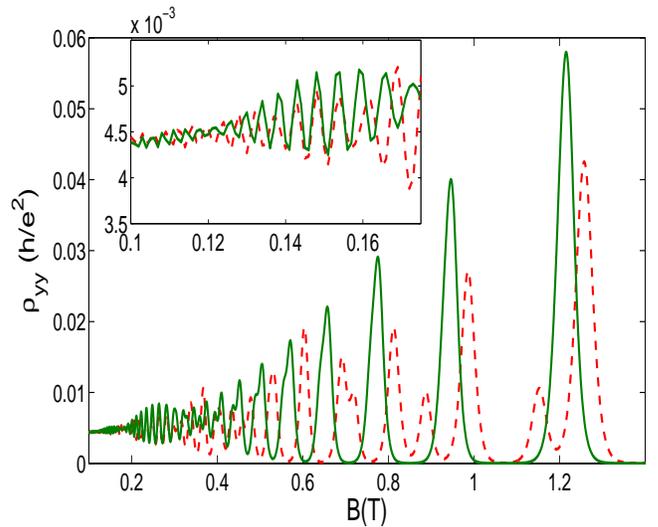}
\caption{(color online) Plot of the collisional resistivity $\rho_{yy}$
as a function of magnetic field $B$ for 
$\alpha=0.04$ eVnm$^3 $ (solid green)
and $\alpha=0.10$ eVnm$^3 $ (dashed red). Inset shows 
the beating patterns in the low magnetic field range.} 
\label{Fig5}
\end{center}
\end{figure}

\begin{figure}[t]
\begin{center}\leavevmode
\includegraphics[width=93mm,height=65mm]{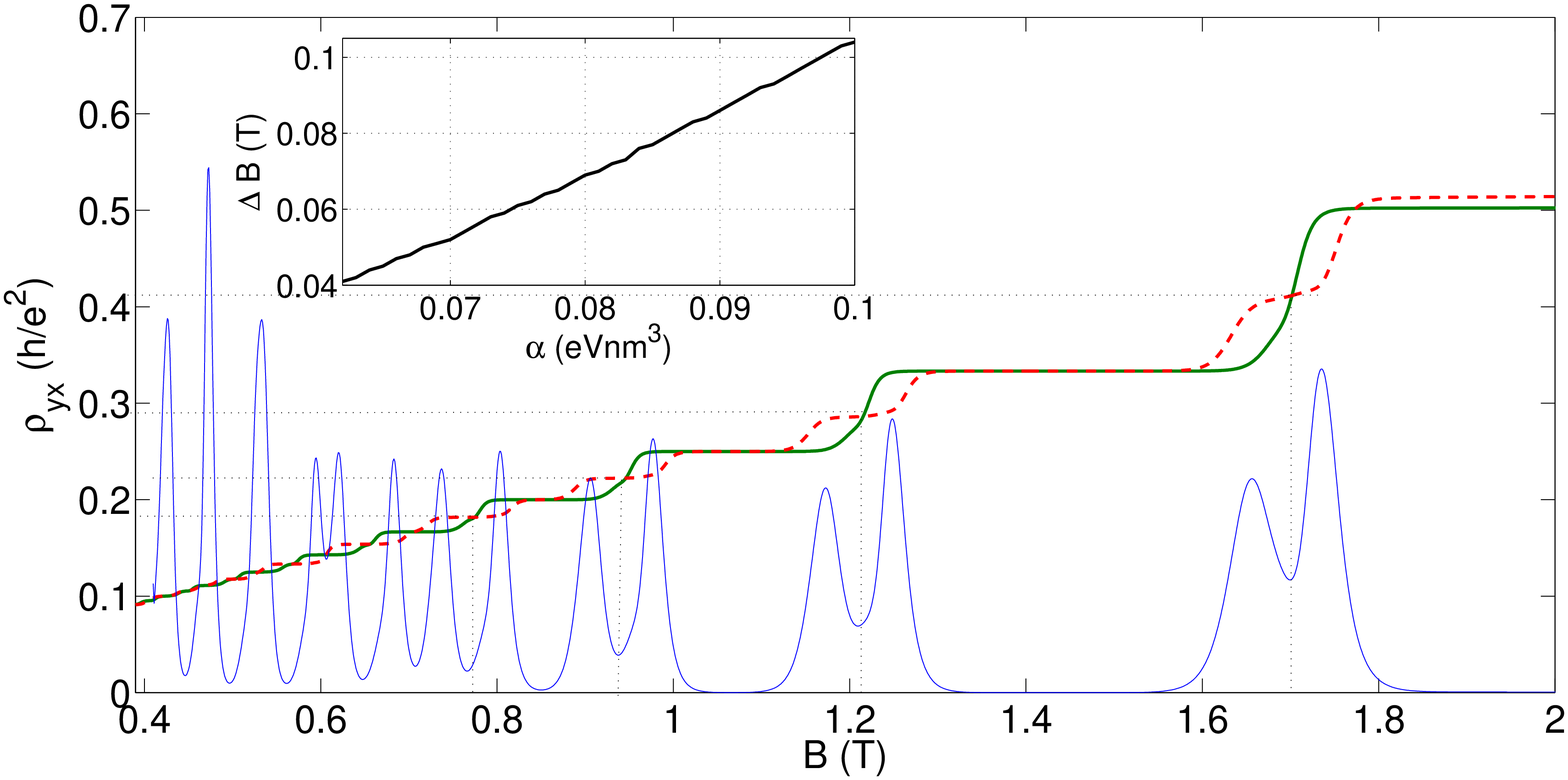}
\caption{(color online) Plot of the Hall resistivity $\rho_{yx}$ vs 
$B$ for  
$\alpha=0.04$ eVnm$^3$ (solid green) and 
and $\alpha=0.10$ eVnm$^3$ (dashed red).
The curve shown in blue thin line is the plot of
$d\rho_{yx}/dB$ for 
$\alpha=0.10$ eVnm$^3$. In the inset the width of the additional plateau 
is plotted vs $\alpha$ around B=1.2T.}
\label{Fig3}
\end{center}
\end{figure}

\begin{table}[ht]
\centering
\caption{Beating nodes calculated from Eq. (\ref{beat}) and
obtained from Fig. 1 are tabulated here.}
\begin{center}
\begin{tabular}{ |c|c|c|c|c| }
\hline
\hline
$\alpha$ & \multicolumn{2}{ |c| }{$0.048$ eVnm$^3$}  
& \multicolumn{2}{ |c| }{$0.080$ eVnm$^3$} \\
\hline
$j$  &  Fig. 1(b)  &  Eq. (\ref{beat}) & Fig. 1(a) & Eq. (\ref{beat}) \\ \hline
1  &  2.730  &  2.375    & 1.710    & 1.425   \\
2  &  4.960  &  4.861    & 3.070    & 2.916   \\
3  &  7.075  &  7.114    & 4.260    & 4.268   \\
4  &  9.295  &  9.305    & 5.550    & 5.583   \\
5  &    -    &    -      & 6.675    & 6.882   \\
6  &    -    &    -      & 8.010    & 8.172   \\
7  &    -    &    -      & 9.340    & 9.458   \\
\hline
\end{tabular}
\end{center}
\end{table}

To see the high magnetic field behavior of the longitudinal resistivity, 
it is convenient to plot $\rho_{yy} $ vs magnetic field $B$.
In Fig. 2 we plot longitudinal resistivity $ \rho_{yy} $ 
versus magnetic field $B$ for different values of $\alpha$.
The height of the peaks in the SdH oscillations reduce with the increase of $\alpha $.
It shows that at low $B$ field ($B < 0.5$ T) 
the regular beating pattern appears in the SdH oscillations. 
At high magnetic field, the SOI effect is reduced and 
the resistivity peaks split into two, instead of showing
the regular beating pattern.

Figure 3 shows the variation of the Hall resistivity $\rho_{yx}$ 
versus magnetic field $B$ for two different values of $\alpha$. 
One can see that the integer quantum Hall plateaus occur at
$h/(e^2N)$, where $N$ is integer.  
It is interesting to note that as $\alpha$ increases there is an 
additional plateau at $h/(e^2(N+1/2))$ appearing between any two conventional 
plateaus. 
On the other hand, one can find from the inset of Fig. 3 that the width of the additional 
plateau increases with increasing $\alpha$.
In Fig. 3 we also plot $d\rho_{yx}/dB$ versus $B$ as shown by thin line.
The sudden jump in the conventional Hall resistivity is characterized 
by the peaks in $d\rho_{yx}/dB$. 
These peaks split into two when additional plateaus appear 
due to Rashba SOI.

\section{Summary}
We have studied quantum magnetotransport coefficients of 
$k$-cubic Rashba spin-orbit coupled two-dimensional fermionic systems. 
Our numerical analysis shows the appearance of beating patterns in
the SdH oscillations. By drawing analogy with the Rashba spin-orbit 
coupled 2DEG at heterostructure, we proposed empirical forms of the oscillation
frequencies of the spin-split fermions. It yields excellent matching
of the locations of the nodes and number of oscillations between any two 
successive nodes obtained from the exact numerical calculations.
On contrary to the complicated expression (see Eq. 6.39 in Ref. [\onlinecite{Winkler}]) 
for determining spin-orbit coupling constant, we have obtained alternative 
and simple expressions (see Eq. (\ref{alpha}) in this article ) to determine it.
The longitudinal resistivity peaks split into two unequal peaks at high
magnetic filed.
We also found additional Hall plateaus in between any two integer quantum 
Hall plateaus. The appearance of additional Hall plateaus is due to the 
spin-orbit interaction.   
The width of this additional plateau increases with the spin-orbit coupling.

\appendix

\section{Energy Spectrum}
Here we shall derive the energy spectrum and the corresponding eigenstates 
of the Hamiltonian $H$ given by Eq. (\ref{hamil2}). 
With the choice of the Landau gauge ${\bf{A}}=(0,xB,0)$,
$k_y $ is a good quantum number since $[H,p_y]=0$. It allows us to
write the wave function as 
$ \psi(x,y) \sim e^{ik_yy} \Phi(x) $.
Now the Hamiltonian $H$ can be expressed as
\begin{eqnarray}\label{rhamil}
H=\begin{pmatrix}
a^\dagger a+\frac{1}{2}-\chi && \sqrt{8}\tilde{\alpha}{a^\dagger}^3    \\
\sqrt{8}\tilde{\alpha}a^3 && a^\dagger a+\frac{1}{2}+\chi
\end{pmatrix}\hbar\omega_c,
\end{eqnarray}
where $a^\dagger = - (i l_c/\sqrt{2}\hbar) \Pi_- $ and 
$a= (i l_c/\sqrt{2} \hbar) \Pi_+$ are the ladder operators
such that $a^\dagger\phi_n
=\sqrt{n+1}\phi_{n+1} $
and $a\phi_n=\sqrt{n}\phi_{n-1}$, respectively.
Here, 
$$
\phi_n (X) = \frac{1}{\sqrt{\sqrt{\pi}l_c 2^n n!}} H_n(X/l_c) e^{-X^2/2l_c^2}
$$
are the harmonic oscillator states with $H_n(X)$ being the Hermite polynomial
of order $n$, $ X = x - x_c $.

As seen from Eq. (\ref{rhamil}), one can take one of the following forms
for the wave function $\Phi(X) $:
\begin{eqnarray}\label{W}
  \Phi(X)&=&
  \begin{pmatrix}
    \phi_n(X)\\
    D\phi_{n-3}(X)
  \end{pmatrix} \textrm{or} \hspace{0.1cm}
\Phi(X) =     
\begin{pmatrix}
     D^{\prime}\phi_n(X)\\
    \phi_{n-3}(X)
  \end{pmatrix}
  \end{eqnarray}
which is valid for $n\geq 3$.

Substituting Eq.(\ref{W}) in the time-independent Schroedinger equation 
$H \Phi(x) = E \Phi(x) $, we get
\begin{equation}\label{p}
\begin{aligned}
  n+1/2-\chi+D\tilde{E}_{n\alpha}=\varepsilon _n\\
  \tilde{E}_{n\alpha}+D(n-3+1/2+\chi)=D\varepsilon_n
\end{aligned}
\end{equation} and
\begin{equation}\label{n}
\begin{aligned}
  D^{\prime} (n+1/2-\chi)+\tilde{E}_{n\alpha}=\varepsilon_nD^{\prime}  \\
    D^{\prime}\tilde{E}_{n\alpha}+n-3+1/2+\chi=\varepsilon_n,
\end{aligned}
\end{equation}
where $\tilde{E}_{n\alpha}=\sqrt{8n(n-1)(n-2)}\tilde{\alpha}$ and
$ \varepsilon_n = E_n/(\hbar \omega_c) $.

By solving either Eq.(\ref{p}) or Eq.(\ref{n}) we get the same energy spectrum
\begin{equation}
E_n^{\lambda}=\hbar\omega_c\Big[n-1+
\lambda\sqrt{\tilde{E}_{n\alpha}^2+\tilde{E}_0^2}\Big],
\end{equation}
where $\lambda=\pm$ and
$\tilde{E}_0=3/2-\chi$.

Putting $ \varepsilon_n^+ =  n-1+\sqrt{\tilde{E}_{n\alpha}^2+\tilde{E}_0^2}$ in Eq.(\ref{p}) and
$\varepsilon_n^- = n-1-\sqrt{\tilde{E}_{n\alpha}^2+\tilde{E}_0^2}$ in Eq.(\ref{n}), we get
\begin{eqnarray}
  D=D_n=\frac{\tilde{E}_{n\alpha}}{\tilde{E}_0 +
\sqrt{\tilde{E}_0^2+\tilde{E}_{n\alpha}^2}}
\end{eqnarray} and
$D^{\prime}=-D=-D_n$.
The normalization factor is
$1/\sqrt{A_n}$ where $A_n=1+D_n^2$. Thus

\begin{eqnarray}
  \Phi_{n}^+(X)&=&\frac{1}{\sqrt{A_n}}
  \begin{pmatrix}
    \phi_n(X)\\
    D_n\phi_{n-3}(X)
  \end{pmatrix}
  \end{eqnarray}
  and
  \begin{eqnarray}
  \Phi_{n}^-(X)&=&\frac{1}{\sqrt{A_n}}
    \begin{pmatrix}
    -D_n\phi_n(X)\\
    \phi_{n-3}(X)
  \end{pmatrix}
  \end{eqnarray}

For $n<3$, there are no spin split states, one can choose
\begin{eqnarray}
\psi_{n,k_y}(x,y)=\frac{e^{ik_yy}}{\sqrt{L_y}}\phi_n
( X )\begin{pmatrix} 1
\\ 0 \end{pmatrix}.
\end{eqnarray}
which will similarly give us
\begin{eqnarray}
  E_n=\big[n+\frac{1}{2}-\chi\big]\hbar\omega_c.
\end{eqnarray}

\section{Form Factors}

The square of the form factors 
$\vert F_{\xi,\xi^\prime}\vert^2$ for $n\geq3$ are given by
\begin{eqnarray}
&&\vert F_{n,n^\prime}^{++}(q)\vert^2=\frac{1}{A_nA_{n^\prime}}
\frac{n^\prime!}{n!}u^{n-n^\prime}
e^{-u}\delta_{k_y^\prime,k_y+q_y}\nonumber\\
&\times&\Bigg[L_{n^\prime}^{n-n^\prime}(u)
+D_nD_{n^\prime}
M_{n,n^\prime}L_{n^\prime-3}^{n-n^\prime}(u)\Bigg]^2
\end{eqnarray}

and

\begin{eqnarray}
 &&\vert F_{n,n^\prime}^{--}(q)\vert^2=\frac{1}{A_nA_{n^\prime}}
\frac{n^\prime!}{n!}u^{n-n^\prime}
 e^{-u}\delta_{k_y^\prime,k_y+q_y}\nonumber\\
&\times&\Bigg[D_nD_{n^\prime}L_{n^\prime}^{n-n^\prime}(u)
+M_{n,n^\prime}L_{n^\prime-3}^{n-n^\prime}(u)\Bigg]^2,
\end{eqnarray}
where $u=q^2l_c^2/2$ and \\ 
$M_{n,n^\prime}=\sqrt{n(n-1)(n-2)/(n^\prime(n^\prime-1)(n^\prime-2))}$.


For $n<3$ we have the following form 
\begin{eqnarray}
\vert F_{nn}\vert^2=e^{-u}L_n^2(u).
\end{eqnarray}

\section{Matrix elements of velocity operator}

Using Heisenberg equation of motion $v_i = (1/i\hbar)[x_i,H]$, we calculate the
following components of the velocity operator
\begin{eqnarray}
v_x=\frac{\Pi_x}{m^\ast}+\frac{3i\alpha}{2\hbar^3}(\sigma_+\Pi_-^2-
\sigma_-\Pi_+^2)
\end{eqnarray}
and
\begin{eqnarray}
v_y=\frac{\Pi_y}{m^\ast}+\frac{3\alpha}{2\hbar^3}(\sigma_+\Pi_-^2+
\sigma_-\Pi_+^2).
\end{eqnarray}
 
The diagonal components of the velocity matrix elements are given by
\begin{eqnarray}\label{vellx}
\langle \zeta,+\vert v_x\vert \zeta^\prime,+\rangle
=ia(B_{n-1}\delta_{n^\prime,n-1}-B_n\delta_{n^\prime,n+1}),
\end{eqnarray}

\begin{eqnarray}
\langle \zeta,-\vert v_x\vert \zeta^\prime,-\rangle
=ia(K_{n-1}\delta_{n^\prime,n-1}-K_n\delta_{n^\prime,n+1}),
\end{eqnarray}

\begin{eqnarray}
\langle \zeta,+\vert v_y\vert \zeta^\prime,+\rangle
=-a(B_n\delta_{n^\prime,n+1}+B_{n-1}\delta_{n^\prime,n-1}),
\end{eqnarray}

and

\begin{eqnarray}
\langle \zeta,-\vert v_y\vert \zeta^\prime,-\rangle
=-a(K_n\delta_{n^\prime,n+1}+K_{n-1}\delta_{n^\prime,n-1}),
\end{eqnarray}

where $a=\omega_cl_c\delta_{k_y^\prime,k_y}$, 
$\vert\zeta\rangle=\vert n,k_y\rangle$
$B_n=(G_n+6\tilde{\alpha}D_{n+1}\sqrt{n(n-1)})/\sqrt{A_nA_{n+1}}$,
$K_n=(F_n-6\tilde{\alpha}D_{n}\sqrt{n(n-1)})/\sqrt{A_nA_{n+1}}$ with 
$G_n=\sqrt{(n+1)/2}+D_nD_{n+1}\sqrt{(n-2)/2}$, and 
$F_n=D_nD_{n+1}\sqrt{(n+1)/2}+\sqrt{(n-2)/2}$.

The off-diagonal components of the velocity matrix elements are 
given by

\begin{eqnarray}
\langle \zeta,+\vert v_x\vert \zeta^\prime,-\rangle
=ia(K_n^\prime\delta_{n^\prime,n+1}-B_{n-1}^\prime\delta_{n^\prime,n-1}),
\end{eqnarray}

\begin{eqnarray}
\langle \zeta,-\vert v_x\vert \zeta^\prime,+\rangle
=ia(B_n^\prime\delta_{n^\prime,n+1}-K_{n-1}^\prime\delta_{n^\prime,n-1}),
\end{eqnarray}

\begin{eqnarray}
\langle \zeta,+\vert v_y\vert \zeta^\prime,-\rangle
=a(K_n^\prime\delta_{n^\prime,n+1}+B_{n-1}^\prime\delta_{n^\prime,n-1}),
\end{eqnarray}

\begin{eqnarray}\label{velly}
\langle \zeta,-\vert v_y\vert \zeta^\prime,+\rangle
=a(B_n^\prime\delta_{n^\prime,n+1}+K_{n-1}^\prime\delta_{n^\prime,n-1}),
\end{eqnarray}
where $B_n^\prime= 
(G_n^\prime+6\tilde{\alpha}D_{n}D_{n+1}\sqrt{n(n-1)})/\sqrt{A_nA_{n+1}}$,
$K_n^\prime=(F_n^\prime-6\tilde{\alpha}\sqrt{n(n-1)})/\sqrt{A_nA_{n+1}}$
with $G_n^\prime=D_n\sqrt{(n+1)/2}-D_{n+1}\sqrt{(n-2)/2}$ and
$F_n^\prime=D_{n+1}\sqrt{(n+1)/2}-D_n\sqrt{(n-2)/2}$.


\begin{thebibliography}{55}




\bibitem{rashba1}
E. I. Rashba, 
Sov.Phys. Solid State {\bf 2}, 1109 (1960).

\bibitem{rashba2}
Y. A. Bychkov and E. I. Rashba, 
J. Phys. C: Solid State Phys. {\bf 17}, 6039 (1984).

\bibitem{alpha1}
J. Nitta, T. Akazaki, H. Takayanagi, and T. Enoki,
Phys. Rev. Lett. {\bf 78}, 1335 (1997).

\bibitem{alpha2}
T. Matsuyama, R. Kursten, C. Meibner, and U. Merkt,
Phys. Rev. B {\bf 61}, 15588 (2000).

\bibitem{alpha3}
S. J. Papadakis, E. P. De Poortere, H. C. Manoharan†, J. B. Yau, 
M. Shayegan, and S. A. Lyon, 
Phys. Rev. B {\bf 65}, 245312 (2002).


\bibitem{datta}
S. Datta and B. Das, 
Appl. Phys. Lett. {\bf 56}, 665 (1990).


\bibitem{cahay}
S. Bandyopadhyay and M. Cahay, Introduction to
Spintronics (CRC press-2008).

\bibitem{sarma}
I. Zutic, J. Fabian, and S. Das Sarma,
Rev. Mod. Phys. {\bf 76}, 323 (2004).

\bibitem{acta}
J. Fabian, A. Matos-Abiague, C. Ertler, P. Stano, and I. Zutic,	
Acta Physica Slovaca {\bf 57}, 565(2007).

\bibitem{2dh1}
R. Winkler,
Phys. Rev. B {\bf 62}, 4245 (2000).

\bibitem{2dh2}
J. Schliemann and D. Loss,
Phys. Rev. B {\bf 71}, 085308 (2005).

\bibitem{2dh3}
B. A. Bernevig and S. C. Zhang,
Phys. Rev. Lett. {\bf 95}, 016801 (2005).

\bibitem{Winkler}
R. Winkler, Spin-Orbit Coupling Effects in 
Two-Dimensional Electron and Hole Systems
(Springer Verlag-2003).


\bibitem{SrTi1}
H. Nakamura, T. Koga, and T. Kimura,
Phys. Rev. Lett. {\bf 108}, 206601 (2012).

\bibitem{SrTi}
Z. Zhong, A. Toth, and K. Held,
Phys. Rev. B {\bf 87}, 161102(R), (2013).

\bibitem{moriya}
R. Moriya et al.,
Phys. Rev. Lett. {\bf 113}, 086601 (2014).


\bibitem{luttin1}
J. M. Luttinger and W. Kohn,
Phys. Rev. {\bf 97}, 869  (1955).

\bibitem{luttin2}
J. M. Luttinger,
Phys. Rev. {\bf 102}, 1030 (1956).




\bibitem{Grbic}
B. Grbic, R. Leturcq, T. Ihn, K. Ensslin, D. Reuter,
and A. D. Wieck, 
Phys. Rev. B {\bf 77}, 125312 (2008).


\bibitem{mass1}
Y. T. Chiu, M. Padmanabhan, T. Gokmen, J. Shabani, E. Tutuc, 
M. Shayegan, and R. Winkler,
Phys. Rev. B {\bf 84}, 155459 (2011).

\bibitem{mass2}
T. M. Lu, Z. F. Li, D. C. Tsui, M. J. Manfra, L. N. Pfeiffer,
and K. W. West, 
Appl. Phys. Lett. {\bf 92}, 012109 (2008). 

\bibitem{mass3}
F. Nichele, A. N. Pal, R. Winkler, C. Gerl, W. Wegscheider,
T. Ihn, and K. Ensslin,
Phys. Rev. B {\bf 89}, 081306 (R) (2014).








\bibitem{lande}
R. Winkler, S. J. Papadakis, E. P. De Poortere, and M. Shayegan,
Phys. Rev. Lett. {\bf 85}, 4574 (2000).


\bibitem{polar}
R. Winkler,
Phys. Rev. B {\bf 71}, 113307 (2005).

\bibitem{rotation}
M. G. Pala, M. Governale, J. Konig and U. Zulicke, and I. Iannaccone, 
Phys. Rev. B {\bf 69}, 045304 (2004).

\bibitem{she1}
S. Murakami, N. Nagaosa, and S. C. Zhang,
Science {\bf 301}, 1348 (2003).

\bibitem{she2}
J. Wunderlich, B. Kaestner, J. Sinova, and T. Jungwirth,
Phys. Rev. Lett. {\bf 94}, 047204 (2005).

\bibitem{she3}
M. W. Wu and J. Zhou,
Phys. Rev. B {\bf 72}, 115333 (2005).

\bibitem{she4}
K. Nomura, J. Wunderlich, J. Sinovo, B. Kaestner, A. H. MacDonald, and T. Jungwirth,
Phys. Rev. B {\bf 72}, 245330 (2005).

\bibitem{she5}
W. Q. Chen, Z. Y. Weng, and D. N. Sheng, 
Phys. Rev. B {\bf 72}, 235315 (2005).

\bibitem{she6}
P. Kleinert and V. V. Bryksin,
Phys. Rev. B {\bf 76}, 073314 (2007).


\bibitem{SrTi2}
L. F. Mattheiss,
Phys. Rev. B {\bf 6}, 4718 (1972).

\bibitem{SrTi3}
R. Bistritzer, G. Khalsa, and A. H. MacDonald,
Phys. Rev. B {\bf 83}, 115114 (2011).

\bibitem{SrTi4}
Z. S. Popovic, S. Satpathy, and R. M. Martin, 
Phys. Rev. Lett. {\bf 101}, 256801 (2008).

\bibitem{SrTi5}
A. F. Santander-Syro et al.,
Nature (London) {\bf 469}, 189 (2011).


\bibitem{Liu}
T. Ma and Q. Liu,
Appl. Phys. Lett. {\bf 89}, 112102 (2006).

\bibitem{Zarea}
M. Zarea and S. E. Ulloa,
Phys. Rev. B {\bf 73}, 165306 (2006).


\bibitem{kubo}
G. M. Eliashberg,
Sov. Phys. JETP {\bf 14(4)}, 866 (1962).
 
\bibitem{Van}
M. Charbonneau, K. M. van Vliet, and P. Vasilopoulos 
J. Math. Phys. {\bf 23}, 318 (1982).

\bibitem{Carol}
P. Vasilopoulos and C. M. Van Vliet, 
J. Math. Phys. {\bf 25}, 1391 (1984).


\bibitem{vasilo}
P. Vasilopoulos,
Phys. Rev. B {\bf 32}, 771 (1985).

\bibitem{Peet}
F. M. Peeters and P. Vasilopoulos,
Phys. Rev. B {\bf 46}, 4667 (1992).



\bibitem{wang}
X. F. Wang and P. Vasilopoulos,
Phys. Rev. B {\bf 67}, 085313 (2003).




\bibitem{Srmass}
A. D. Caviglia, S. Gariglio, C. Cancellieri, B. Sacepe,
A. Fete, N. Reyren, M. Gabay, A. F. Morpurgo, and J.M. Triscone, 
Phys. Rev. Lett. {\bf 105}, 236802 (2010).


\bibitem{fogler}
M. M. Fogler and Shklovskii,
Phys. Rev. B {\bf 52}, 17366 (1995).


\bibitem{Yuan}
Z. Q. Yuan, R. R. Du, M. J. Manfra, L. N. Pfeiffer, 
and K. W. West, 
Appl. Phys. Lett. {\bf 94}, 052103 (2009).


\bibitem{Firoz}
SK F. Islam and T. K. Ghosh,
J. Phys.: Condens. Matter. {\bf 24}, 035302 (2012).




\end{thebibliography}
\end{document}